\documentclass[prd,twocolumn,nofootinbib]{revtex4}

\usepackage[utf8]{inputenc} 
\usepackage[T1]{fontenc}    
\usepackage{hyperref}       
\usepackage{url}            
\usepackage{amsfonts}       
\usepackage{nicefrac}       
\usepackage{microtype}      
\usepackage{graphicx}       
\graphicspath{{media/}}     
\usepackage{amsmath}
\usepackage{feynmp-auto}
\usepackage{amssymb}
\usepackage{ragged2e}
\usepackage{blindtext}
\usepackage{float}
\usepackage{placeins}
\usepackage{float}
\usepackage{xcolor}

\begin{document}

\title{Constraints for scalars and pseudoscalars from $\left(g-2\right)_l$ and existing $e^+e^-$ colliders}

\author{Aleksandr Pustyntsev}
\affiliation{Institut f\"ur Kernphysik and $\text{PRISMA}^+$ Cluster of Excellence, Johannes Gutenberg Universit\"at, D-55099 Mainz, Germany}
\author{Marc Vanderhaeghen}
\affiliation{Institut f\"ur Kernphysik and $\text{PRISMA}^+$ Cluster of Excellence, Johannes Gutenberg Universit\"at, D-55099 Mainz, Germany}

\date{\today}

\begin{abstract}
Scalars and pseudoscalars with masses in the MeV to GeV range are of interest in different extensions of the Standard Model. Such particles are often associated with dark matter, the strong CP problem and the $\left(g-2\right)_{\mu}$ anomaly. In this work we investigate limits for masses of such particles and their couplings to photons and leptons which can be derived from present and currently operating $e^+e^-$ collider experiments and recent $\left(g-2\right)_{l}$ measurements. Our work expands upon previous studies in several ways, demonstrating that the interplay of both couplings is a decisive factor in this type of analyses.
\end{abstract}

\keywords{Axions and ALPs  \and $e^+e^-$ experiments}

\maketitle

\section{Introduction}\label{sec1}

After being first proposed more than fourty years ago, the Peccei-Quinn mechanism so far remains the most persuasive solution to the strong CP problem, that is, the experimentally observed absence of CP violations in quantum chromodynamics~\cite{Peccei:1977hh, Peccei:1977ur,Weinberg:1977ma, Wilczek:1977pj}. The Peccei-Quinn theory comes down to the postulate of a new global symmetry $U\left(1\right)$, the spontaneous breaking of which gives rise to a new pseudoscalar particle called axion, dynamically driving the CP-violating phase to zero.

Axion-like particles (or ALPs) are a natural generalization of this idea, but are not constrained by a linear mass-coupling relation and are not necessarily associated with the strong CP problem. ALPs arise naturally in various extensions of the Standard Model~\cite{Jaeckel:2010ni,Baker:2013zta,Daido:2017wwb,Bagger:1994hh,Arvanitaki:2009fg} and, in addition, both the QCD axion and ALPs are considered to be promising dark matter candidates as being both very long-lived and weakly-interacting~\cite{Daido:2017wwb,Adams:2022pbo,Ringwald:2012hr,Arias:2012az, Giannotti:2015kwo}. 

Over the past decades, extensive efforts have been made to search for axions and ALPs, encompassing both laboratory searches and astronomical observations~\cite{Graham:2015ouw, Irastorza:2018dyq,Bauer:2017ris,Mimasu:2014nea,Safronova:2017xyt,Izaguirre:2016dfi,Regis:2020fhw,BaBar:2021ich}. Recently, there has been a surge of interest in the MeV to GeV range~\cite{Dolan:2017osp,Alves:2017avw,Millea:2015qra,Jaeckel:2015jla,Beacham:2019nyx,Gavela:2019cmq,Mimasu:2014nea,Bauer:2017ris,Agrawal:2021dbo,Antel:2023hkf}. This area remains relatively unexplored by current experiments, particularly when contrasted with the sub-MeV mass region where astrophysical constraints can be applied. Additionally, most of the existing studies assume a single ALP-photon coupling scenario, while the recent works~\cite{Liu:2023bby,Liu:2022tqn,Pustyntsev:2023rns} indicate that accounting for both ALP-photon and ALP-lepton couplings is important in deducing empirical constraints from processes involving leptons.

Meanwhile, the measurement of the muon's anomalous magnetic moment $\left(g-2\right)_{\mu}$ is a powerful tool to test potential effects beyond the Standard Model due to its enhanced sensitivity to New Physics~\cite{Aoyama:2020ynm,Athron:2021iuf}. With the current Standard Model calculations of $\left(g-2\right)_{\mu}$~\cite{Aoyama:2020ynm}, there is an unexplained up to $5 \sigma$ discrepancy\footnote[1]{The recent lattice calculations by the BMW Collaboration, despite showing much smaller deviation~\cite{Borsanyi:2020mff} from the measured $\left(g-2\right)_{\mu}$ value, exhibit tensions with the analyses based on dispersive methods, which rely on $e^+e^- \to \text{hadrons}$ data. Smaller discrepancy is also favored by the CMD-3 measurements of the $e^+e^- \to \pi^+ \pi^-$ cross section~\cite{CMD-3:2023alj}, but they conflict with other experimental results.} between theory and experiment, namely $a_{\text{exp.}} -\, a_{\text{th.}} = 249\left(48\right) \times 10^{-11}$~\cite{Muong-2:2023cdq}. In this context, scalars and pseudoscalars motivated by the Standard Model extensions could serve as a “two birds with one stone” solution. Nevertheless, it is one of the very few viable one-field explanations of the observed anomaly~\cite{Marciano:2016yhf,Buen-Abad:2021fwq,Liu:2022tqn,Chen:2015vqy,Liu:2018xkx}. 

In the present work both analyses are extended. Focusing on $3\gamma$ final states, we rigorously examine constraints which can be derived from existing $e^+e^-$ collider data for scalar and pseudoscalar particles when lepton and photon couplings are simultaneously taken into consideration. Additionally, we show projections for the upcoming Belle II data collection, which by 2030 is expected to result in a record 50 $\text{ab}^{-1}$ of integrated luminosity, with the prospect of significantly narrowing down the parameter space for potential ALPs in the MeV to GeV mass range, where current constraints remain relatively loose. We combine these bounds with ones which can be derived from the present $\left(g-2\right)_e$ data~\cite{Fan:2022eto,Aoyama:2014sxa} and test the available parameter space for the possible $\left(g-2\right)_{\mu}$ resolution. Our results show that the interplay of lepton and gauge boson ALP couplings plays a crucial role in both studies. 

The paper is organized as follows. Section \ref{sec2} provides the overview of the ALP formalism with the discussion of couplings, potential contribution to $\left(g-2\right)_l$ and $e^+e^-$ annihilation with the corresponding constraints derived. In the section \ref{sec3} we repeat our calculations for a potential scalar New Physics particle. Section \ref{sec4} summarizes our work.

\section{Studying ALPs through $\left(g-2\right)_l$ and $e^+e^-$ colliders}\label{sec2}

In this study, we assume that ALPs in the MeV to GeV range interact with electroweak gauge bosons and leptons, i.e. decay exclusively into visible states and are isolated from the QCD sector. The latter statement is justified by the fact that such couplings typically lead to tightly constrained flavour-changing processes~\cite{Dolan:2014ska,BaBar:2011kau}.

The following section provides a short review of the relevant ALP interactions. Lepton universality implies that the parameter space includes only three variables: the ALP mass $m_a$ and its couplings to photons and electrons, which are denoted by $g_{a \gamma \gamma}$ and $g_{aee}$, respectively. We do not focus on any specific UV-complete ALP model, thus the exact origin of $g_{a \gamma \gamma}$ and $g_{aee}$ is unspecified and these quantities must be treated as effective couplings, with both tree-level values and possible loop corrections to them.

\begin{figure}
\centering
\includegraphics{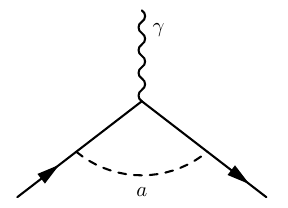} 
\caption{Yukawa-like correction to the lepton dipole moment.}
\label{fig:dyuk}
\end{figure}

In this formalism we provide the calculations of the 1-loop corrections to the $\left(g-2\right)_l$ and of the $e^+e^- \to 3\gamma$ cross section with an intermediate ALP. The signal over background ratio is estimated and constraints are derived for the scenario involving an interplay of both lepton and photon ALP-couplings.

\subsection{Lepton coupling and shift symmetry}\label{subsec:21}

The shift symmetry for an ALP implies that only derivative couplings with leptons are allowed. The lowest order operator of such an effective interaction with the shift symmetry has the form

\begin{equation}\label{eq:der}
\mathcal{L}_{all} = -\frac{g_{all}}{2m_l} \partial_{\mu} a \, \Bar{l} \gamma^5 \gamma^{\mu} l,
\end{equation}
where $l$ stands for the lepton field, $m_l$ denotes its mass and $g_{all}$ is the dimensionless coupling constant. At the level of tree diagram calculations $\mathcal{L}_{all}$ can be equivalently reduced to a pseudoscalar Yukawa interaction 

\begin{equation}\label{eq:yuk}
\mathcal{L}_{all}= -ig_{all} a \, \Bar{l} \gamma^5 l.
\end{equation}

It is clear that lepton universality requires a large enhancement of the ALP coupling to the muon, as compared to the electron, namely

\begin{equation}
    g_{a \mu \mu} \approx \frac{m_{\mu}}{m_e}  g_{a e e},
\end{equation}
and even more enhanced coupling - to the tau-lepton. 

Such an enhanced coupling of ALPs to muons, compared to the one to electrons, opens the possibility of the $\left(g-2\right)_{\mu}$ resolution without the violation of $\left(g-2\right)_e$. Conversely, without this enhanced coupling, resolving the muon anomaly would inevitably conflict with constraints on the electron’s anomalous magnetic moment. 

\begin{figure}
\centering
\includegraphics{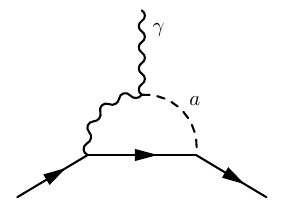} 
\caption{Barr-Zee correction to the lepton dipole moment.}
\label{fig:dbrz}
\end{figure}

\begin{figure*}
\centering
\includegraphics[width=\textwidth]{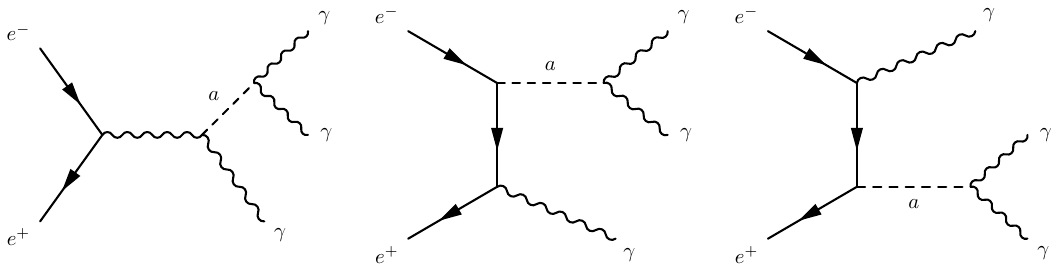} 
\caption{$e^+e^-$ annihilation into three photons involving the $g_{all}$ and $g_{a\gamma\gamma}$ couplings. Graphs obtained from these by crossing are not shown, but are evaluated too.}
\label{fig:tri}
\end{figure*}

The latter already exhibits a relatively strong agreement between theoretical predictions and experimental results. Recent measurements of $\left(g-2\right)_e$ have improved the precision of previous results by a factor of 2.2, achieving an impressive agreement up to 12 digits~\cite{Fan:2022eto}. 

The Yukawa-like 1-loop contribution to the lepton dipole moment is shown in Fig. \ref{fig:dyuk}. It is insensitive to whether \eqref{eq:der} or \eqref{eq:yuk} are used and results in

\begin{equation}
\Delta a_l^Y = - \frac{g_{all}^2}{8\pi^2} \frac{1}{r_a^2} \int_0^1 \frac{\left(1-z\right)^3 }{\frac{\left(1-z\right)^2}{r_a^2}+z}dz,
\end{equation}
where we denoted

\begin{equation}
a_{l} = \frac{\left(g-2\right)_l}{2}, \quad \quad \quad r_a=\frac{m_a}{m_l}.
\end{equation}

As this contribution is negative, while the $\left(g-2\right)_{\mu}$ discrepancy is positive, it is evident that such a photophobic ALP scenario is not capable of resolving the mentioned anomaly on its own. However, accounting for the additional correction from the ALP-photon interaction, shown in Fig. \ref{fig:dbrz}, changes the situation, since this Barr-Zee contribution can be of an opposite sign. Consequently, the sum of the diagrams shown in Figs. \ref{fig:dyuk} and \ref{fig:dbrz} can yield a positive overall contribution to $a_{l}$.

\subsection{Low energy interactions between ALPs and gauge bosons}\label{subsec:22}

The generic gauge-invariant Lagrangian of ALPs interaction with electroweak vector bosons is given by~\cite{Dolan:2017osp}

\begin{equation}\label{eq:ew}
\mathcal{L}_{aEW} = -\frac{g_{aBB}}{4}aB^{\mu \nu}\tilde{B}_{\mu \nu} -\frac{g_{aWW}}{4}a \textbf{W}^{\mu \nu}\tilde{\textbf{W}}_{\mu \nu},
\end{equation}
$B^{\mu \nu}$ and $\textbf{W}^{\mu \nu}$ refer to the $U\left(1\right)$ and $SU\left(2\right)$ field tensors, respectively. The corresponding dual pseudotensors are defined in the standard way, $\tilde{B}_{\mu \nu} = \frac{1}{2}\varepsilon_{\mu \nu \lambda \sigma} {B}^{ \lambda \sigma}$ and similarly for $\textbf{W}^{\mu \nu}$, while $g_{a BB}$ and $g_{a WW}$ are the coupling constants of $\mbox{GeV}^{-1}$ dimension.

After the symmetry breaking the effective interaction of ALPs and photons is given by

\begin{equation}
\mathcal{L}_{a \gamma \gamma} = -\frac{g_{a \gamma \gamma}}{4}aF^{\mu \nu}\tilde{F}_{\mu \nu} - \frac{g_{a \gamma Z}}{2}aF^{\mu \nu}\tilde{Z}_{\mu \nu},
\end{equation}
where the new coupling constants were defined

\begin{align}
& g_{a\gamma\gamma} = g_{aBB} \cos^2{\theta_w}+g_{aWW} \sin^2{\theta_w}, \\
& g_{a\gamma Z} = \frac{g_{aWW}-g_{aBB}}{2} \sin{\left(2\theta_w\right)},
\end{align}
with $\theta_w$ being the weak mixing angle. The ratio of $g_{aBB}$ and $g_{aWW}$ is of key importance for ALPs studies at $e^+e^-$ colliders. In particular, $a\gamma Z$ interaction potentially leads to the anomalous decay $Z \to 3\gamma$, which was a subject of LEP studies~\cite{L3:1994shn,L3:1995nbq}. Such a process would be suppressed if the condition $g_{aWW} \approx g_{aBB}$ is fulfilled.

Searches for flavor-changing processes, however, impose stringent constraints on the coupling $g_{aWW}$~\cite{Izaguirre:2016dfi,BaBar:2021ich}, while the limits on $g_{aBB}$ are more relaxed. It is therefore reasonable to assume that the coupling between ALPs and $\textbf{W}$ is, at best, subdominant, i.e. $g_{aBB} \gg g_{aWW}$. In this case one can express

\begin{equation}\label{eq:ratio}
g_{a\gamma\gamma} \approx -\cot{\theta_w} g_{a\gamma Z} \approx -1.9 \, g_{a\gamma Z}.
\end{equation}

Of course, the Lagrangian \eqref{eq:ew} also leads to a non-zero $aZ Z$ and  $aWW$ interactions. However, they do not contribute to $\left(g-2\right)_l$ at 1-loop level and are irrelevant in the domain of interest corresponding to $e^+e^-$ colliders with center-of-momentum energy below 10 GeV, since the direct production of $Z$ and $W$ bosons is not possible. 

The corresponding correction to $\left(g-2\right)_l$ is shown in Fig. \ref{fig:dbrz} with the internal boson line standing for either photon or $Z$. The leading (log-enhanced) contribution is

\begin{equation}\label{eq:abz}
\begin{split}
& \Delta a_l^{BZ} \simeq \\ 
& \frac{ g_{all} m_l}{8\pi^2}\ln{\Lambda^2} \left(g_{a\gamma\gamma}-\frac{4\sin^2{\theta_w}-1}{4\sin{\theta_w}\cos{\theta_w}}g_{a\gamma Z}\right),
\end{split}
\end{equation}
where $\Lambda$ denotes the high-energy cut-off, referring to the new physics scale, which we set to $1$ TeV.   

Clearly, the effect of $g_{a\gamma Z}$ interaction is suppressed and provides only a small correction to $g_{a\gamma \gamma}$ contribution. The expressions for finite parts are summarized in Appendix \ref{appendix:a}.

It is useful to note that the overall sign of \eqref{eq:abz} depends on the relative sign between $g_{a\gamma\gamma}$ and $g_{all}$. We assume them to be of the same sign - in this case $\Delta a_l^{BZ}$ is positive, which is consistent with $\left(g-2\right)_{\mu}$ anomaly.

Finally, we note that in the Lagrangian \eqref{eq:ew}, we restricted ourselves to CP-even interactions, which do not contribute to electric dipole moments (EDMs) at 1-loop level. CP-odd ALPs have been studied in, e.g. \cite{Kirpichnikov:2020lws,Choi:2023bou,Marciano:2016yhf}, and are tightly constrained by the electron EDM measurements\footnote[2]{Electrophobic ALPs, which couple exclusively to heavy leptons, could, in principle, evade this constraint, but we do not discuss such a specific scenario here.}~\cite{Roussy:2022cmp}. 

\subsection{ALP searches at $e^+e^-$ colliders}\label{subsec:23}

One of the most promising methods for detecting ALPs is by probing their coupling to photons. Numerous experiments have been dedicated to this search over the past years, targeting ALPs in cosmic rays~\cite{Graham:2015ouw, Irastorza:2018dyq}, atomic and molecular effects~\cite{Safronova:2017xyt}, beam dumps and collider experiments~\cite{Bauer:2017ris,Mimasu:2014nea,BESIII:2022rzz,Belle-II:2020jti,CMS:2018erd,Blumlein:1990ay,NA64:2020qwq,PhysRevLett.59.755,BERGSMA1985458,Jaeckel:2015jla,ATLAS:2020hii,Knapen:2016moh}. A summary of current limits on ALP couplings can be found in~\cite{AxionLimits}. 

It this work, we focus on $e^+e^-$ collider experiments, which can probe potential ALPs in the mass range $100 \, \text{MeV} \lesssim m_a \lesssim 10 \, \text{GeV}$. Lower mass particles can be probed with beam dump experiments, such as NA64~\cite{NA64:2020qwq,NA64:2021aiq,Dusaev:2020gxi}, while heavier candidates are subject of LHC studies~\cite{CMS:2018erd,ATLAS:2020hii,Knapen:2016moh}.

Currently, there are results from a Belle II 2018 test run with the 445 $\text{pb}^{-1}$ of integrated luminosity. The data allowed to establish an upper limit on the cross section $\sigma_a$ for the process $e^+e^- \to 3\gamma$ involving an intermediate in the mass range $0.2 < m_a < 9.7$ GeV~\cite{Belle-II:2020jti}. The corresponding coupling limit was derived under the assumption that $a \to 2 \gamma$ is the only allowed decay channel, making use of the formula

\begin{figure}
\centering
\includegraphics[width=0.4\textwidth]{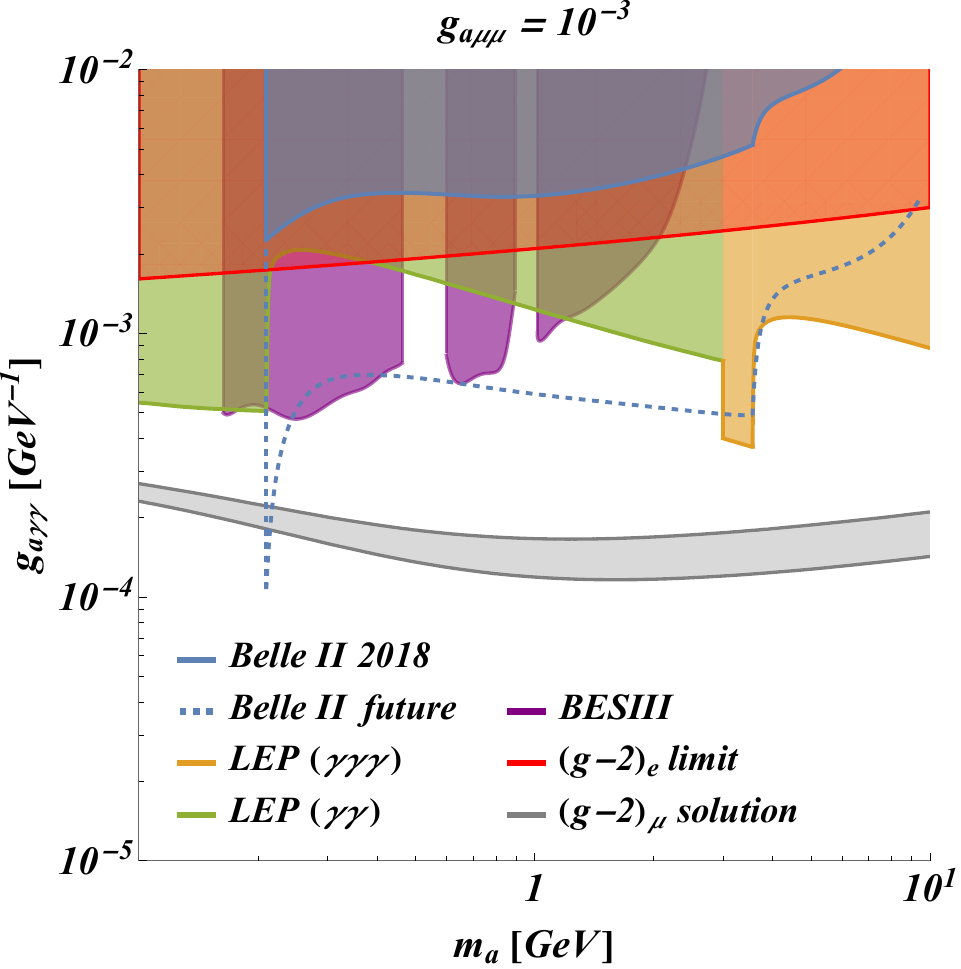} 
\caption{Exclusion limits on the $\left(m_a, g_{a\gamma\gamma}\right)$ parameter space for the fixed value of $g_{a\mu\mu} = 10^{-3}$. The colored regions represent the existing bounds, the blue dashed line is the projection for the upcoming data collection at Belle II. The grey band refers to the parameter space that could potentially resolve the $\left(g-2\right)_{\mu}$ discrepancy.}
\label{fig:agmumu}
\end{figure}

\begin{equation}
\left.\sigma_{a}\right|_{g_{all}=0} = \frac{\alpha g^2_{a \gamma \gamma}}{24} \left(1 - \frac{m_a^2}{s} \right)^3,
\end{equation}
where $\alpha \equiv e^2/4\pi$. The integration over the full solid angle $4\pi$ is implied, but due to the absence of forward-backward enhancement in this process, it leads only to a minor correction. The outcomes of this test run were already on par with the results from the LEP run at the $Z$-pole, which we discuss below. 

Projections for the upcoming Belle II data collection were also studied~\cite{Dolan:2017osp} and later generalized to the scenario when ALP-electron interaction is taken into consideration, leading to a notable modification~\cite{Pustyntsev:2023rns,Liu:2023bby}. Fig. \ref{fig:tri} shows the relevant diagrams for such search. The total decay width of ALPs is assumed to be a sum of the following contributions

\begin{figure}
\centering
\includegraphics[width=0.4\textwidth]{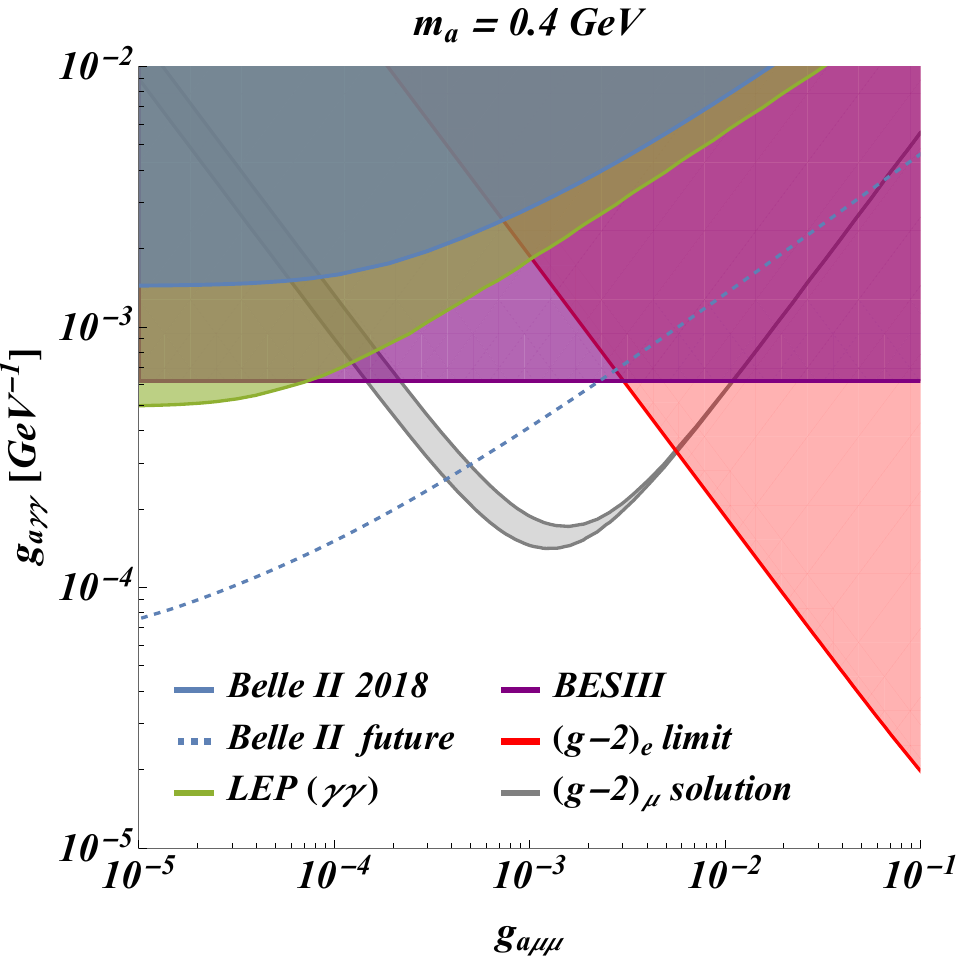} \\
\textcolor{white}{1}\\ 
\includegraphics[width=0.4\textwidth]{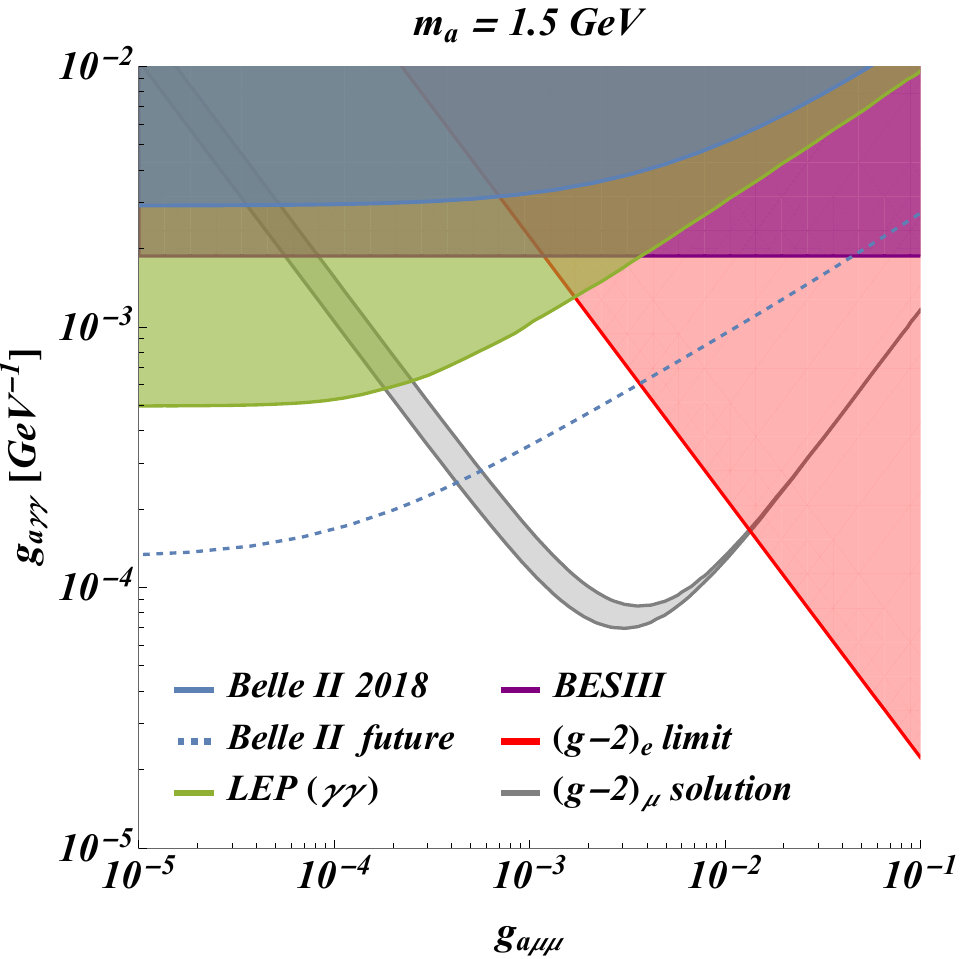} \\
\textcolor{white}{1}\\ 
\includegraphics[width=0.4\textwidth]{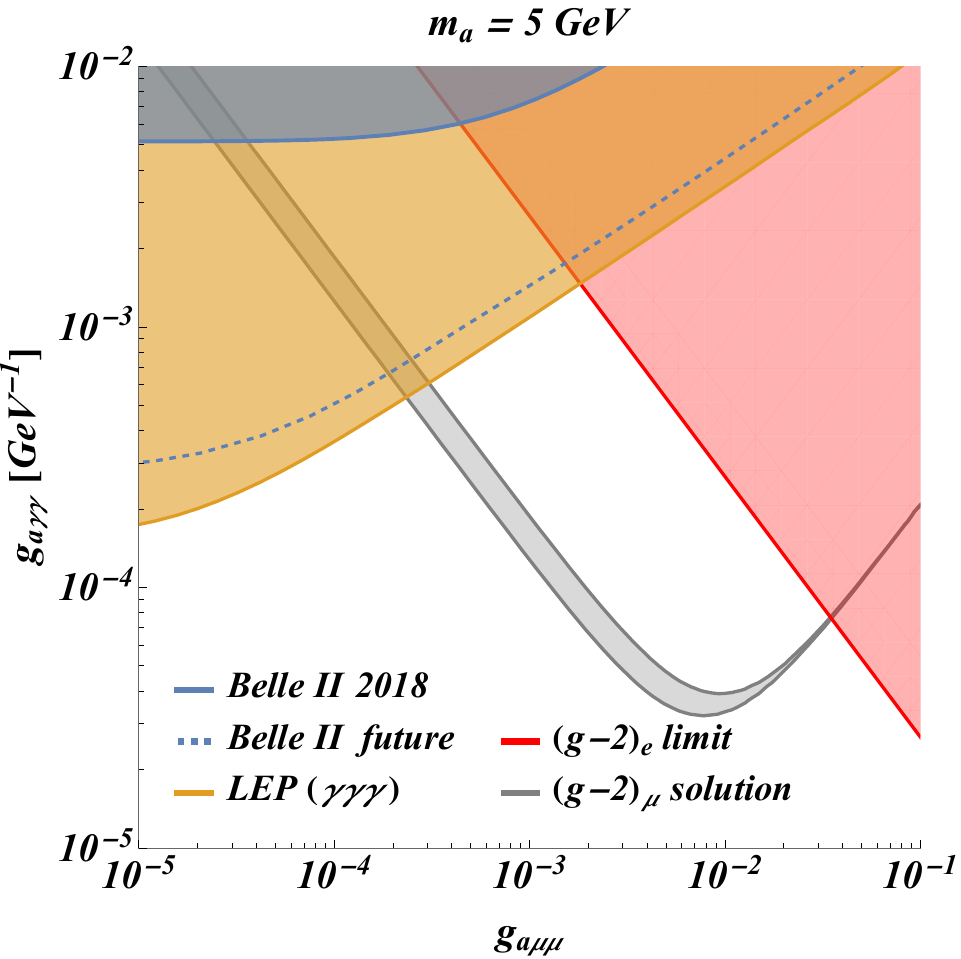} 
\caption{Exclusion limits on the $\left(g_{a\mu\mu}, g_{a\gamma\gamma}\right)$ parameter space for masses $m_a = 0.4$, $1.5$ and $5$ GeV. The colored regions represent the existing bounds, the blue dashed line is the projection for the upcoming data collection at Belle II. The grey band refers to the parameter space that could potentially resolve the $\left(g-2\right)_{\mu}$ discrepancy.}
\label{fig:alps}
\end{figure}

\begin{equation}
\Gamma_a = \Gamma_{a\gamma\gamma} +  \sum_{e,\mu,\tau} \Gamma_{all},
\end{equation}
where we denoted

\begin{align}
& \Gamma_{a \gamma \gamma} = \frac{g^2_{a \gamma \gamma}m_a^3}{64\pi} ,\\
& \Gamma_{all} = \frac{g_{all}^2}{4\pi}  \sqrt{\frac{m^2_a}{4}-m_{l}^2} \,  \theta\left[m_a - 2m_l\right].
\end{align}

For comparable values of $m_a g_{a\gamma\gamma}$ and $g_{a\mu\mu}$ it is typically dominated by leptonic decays.

To determine the exclusion limits, we integrate the cross section over the Belle II geometry and require three resolved photons within the $37.3^{\circ} <\theta< 123.7^{\circ}$ angular region in the lab frame (same as in the 2018 analysis). The event selection threshold is set to $1$ GeV in the center-of-momentum frame for the 2018 data and $0.25$ GeV for the projection. For a gaussian distributed signal the aimed sensitivity can be estimated as

\begin{equation}
\frac{\sigma_{a}}{\sigma_{b}} = \frac{N}{\sqrt{L \cdot \sigma_{b}}},
\end{equation}
where $L$ denotes the integrated luminosity, $N$ is the number of standard deviations that determines whether or not a fluctuation is considered as a signal and $\sigma_{b}$ is the background cross section (which is primarily dominated by the QED $e^+e^- \to 3\gamma$ annihilation). For our analysis, we set $N = 2$, corresponding to a 95\% confidence level.

In addition to the Belle II exclusion limits, we also incorporate the adapted boundaries from the results of LEP run at $Z$-pole~\cite{L3:1995nbq,L3:1994shn}, which imposed stringent constraints on possible decays $Z\to 2\gamma$ and $Z\to 3\gamma$, significantly narrowing down potential New Physics parameter space. Furthermore, we include results from BESIII~\cite{BESIII:2022rzz} analyses of $J/ \psi \to 3\gamma$ decays (notably, its results remain independent of $g_{all}$). 

Coupling of potential new particles to leptons, primarly muons, was also a subject of studies at $e^+e^-$ colliders~\cite{BaBar:2020jma,BaBar:2016sci}. BaBar analysis of $\tau^+\tau^- e^+ e^-$ and $\tau^+\tau^- \mu^+ \mu^-$ final states established a boundary $g_{a\mu\mu} \lesssim 10^{-3}$ for ALPs lighter than $4$ GeV (assuming 100$\%$ branching ratio to leptons). The resolution of $\left(g-2\right)_{\mu}$ problem is achieved if the product of couplings satisfy $g_{a\gamma\gamma}g_{a\mu\mu} \approx 10^{-7}$, which then implies $g_{a\gamma\gamma} \approx 10^{-4} \, \text{GeV}^{-1}$.

Finally, let us emphasize that we only considered the scenario of ALP visible decays. However, if ALPs predominantly decay invisibly (e.g., into dark matter particles), the limits from visible decay searches may be inapplicable. Such a possibility was examined in~\cite{Darme:2020sjf} for a broad range of ALP masses, with the strongest constraints in the $100 \, \text{MeV} \lesssim m_a \lesssim 10 \, \text{GeV}$ range provided by the Belle II search for missing energy.

\subsection{Results for pseudoscalars}\label{subsec:24}

For illustrative purposes, we choose the value $g_{a\mu\mu} = 10^{-3}$ to project onto the $\left(m_a,g_{a\gamma\gamma}\right)$ plane in Fig. \ref{fig:agmumu}. It is evident that in the coming years, Belle II will be capable to cover a considerable part of the parameter space for potential New Physics, especially in the lower mass region, where it has a much enhanced sensitivity compared to LEP. 

We would also like to emphasize that the projection curve is quite conservative. The setup and selection procedures are likely to be further optimized compared to the 2018 test run, so we expect the final curve to be even more improved. In particular, upgrading the spatial resolution of photons would allow to access the currently unexplored region $m_a < 0.2 \, \text{GeV}$~\cite{Belle-II:2020jti}.

On the other hand, it is evident that more data probing the ALP-lepton coupling is necessary to comprehensively explore the parameter space relevant to $\left(g-2\right)_{\mu}$. In this work we are focused on $3\gamma$ final states, but the detailed study of leptonic final states at  BESIII or Belle II has the potential to improve upon existing constraints, in the latter case - even with the relatively limited data available from its test run.

Next, fig. \ref{fig:alps} shows our results for three different ALP masses, $m_a = 0.4$, $1.5$ and $5$ GeV as a projection onto the $\left(g_{a\mu\mu},g_{a\gamma\gamma}\right)$ plane. In the limit $g_{all} \to 0$ these bounds are in agreement with the single $g_{a\gamma\gamma}$ coupling scenario. 

\begin{figure}
\centering
\includegraphics[width=0.4\textwidth]{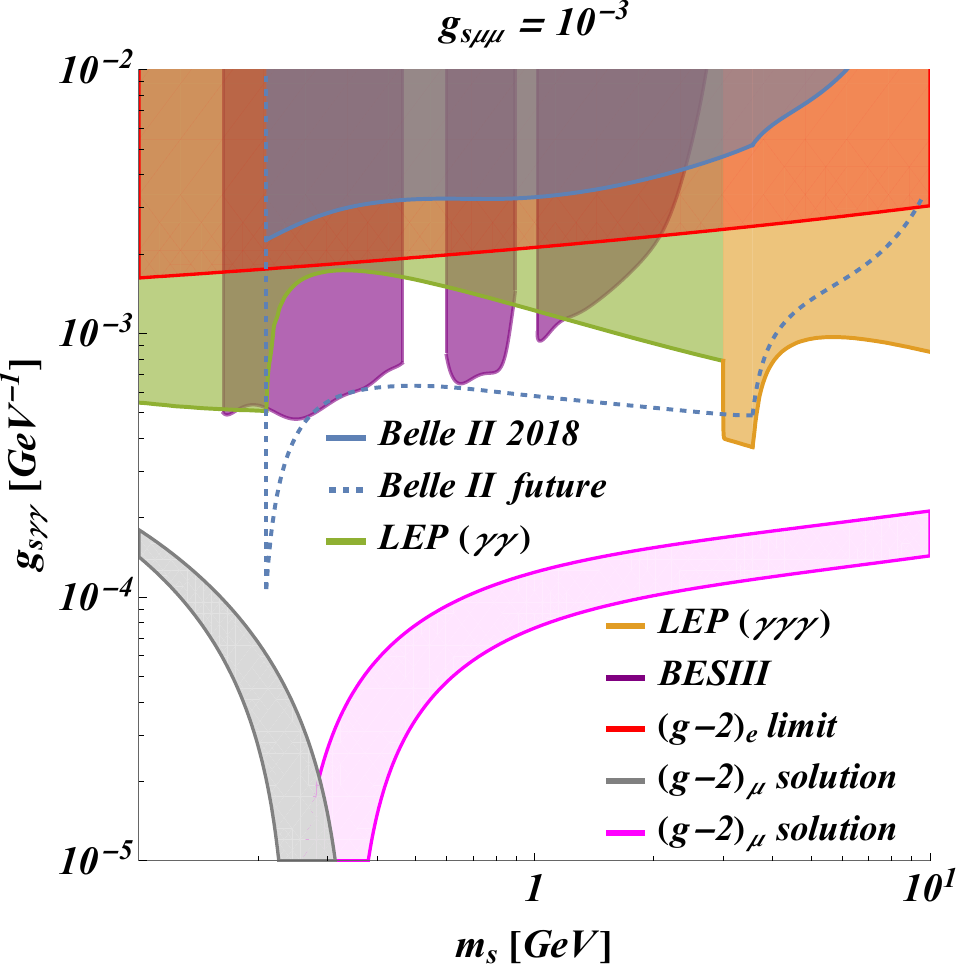} 
\caption{Exclusion limits on the $\left(m_s, g_{s\gamma\gamma}\right)$ parameter space for the fixed value of $g_{s\mu\mu} = 10^{-3}$. The colored regions represent the existing bounds, the blue dashed line is the projection for the upcoming data collection at Belle II. The magenta and grey bands refer to the parameter space that could potentially resolve the $\left(g-2\right)_{\mu}$ discrepancy under the scenarios where $g_{s\gamma\gamma}$ and $g_{s\mu\mu}$ have the same and opposite signs, respectively.}
\label{fig:sgmumu}
\end{figure}

We see that in fact a large part of the parameter space for $\left(g-2\right)_{\mu}$-solving ALPs can be ruled out with experimental data which are already available. A sizeable further part of the parameter space corresponding to the solution for $\left(g-2\right)_{\mu}$ can be tested after Belle II finishes the data collection and BESIII collects more data on $J/ \psi$ decays (at the time of writing this article BESIII collaboration released a pre-print on updated ALP limits~\cite{BESIII:2024hdv}, with a four-fold increased statistics of $J/ \psi$ decays compared to the previous result~\cite{BESIII:2022rzz}).

\begin{figure}
\centering
\includegraphics[width=0.4\textwidth]{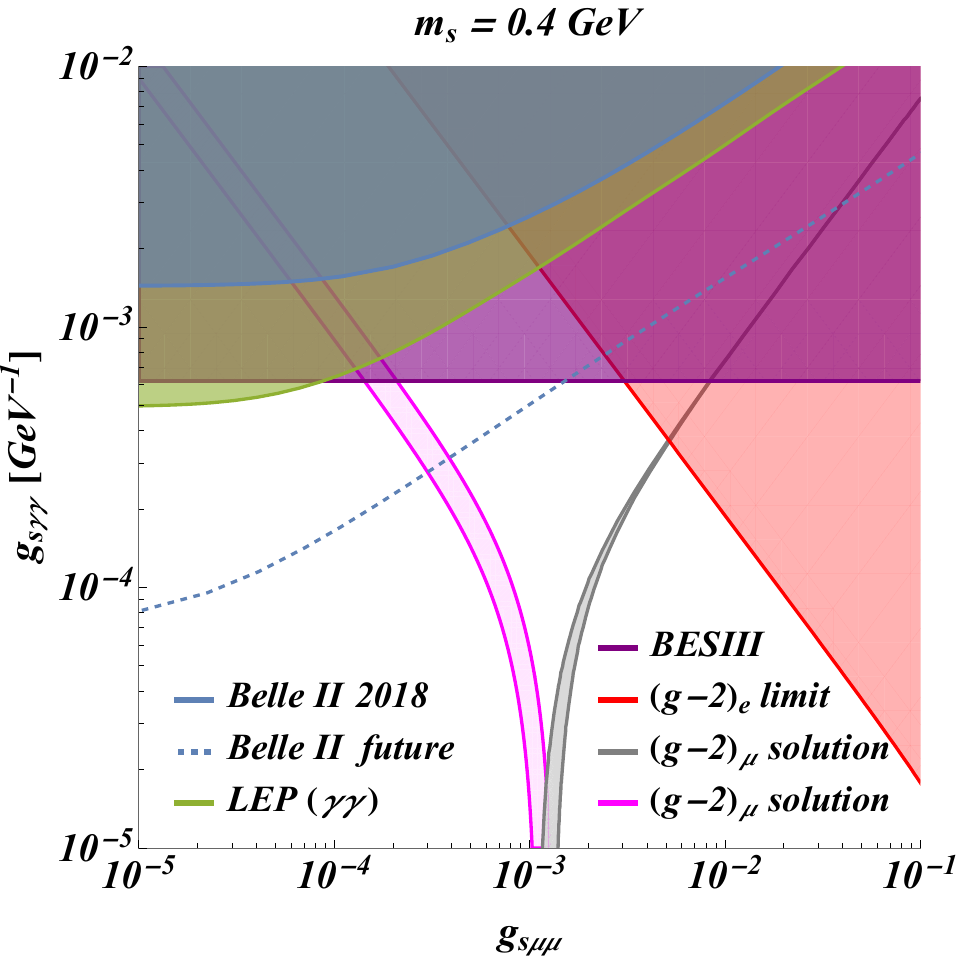} \\
\textcolor{white}{1}\\ 
\includegraphics[width=0.4\textwidth]{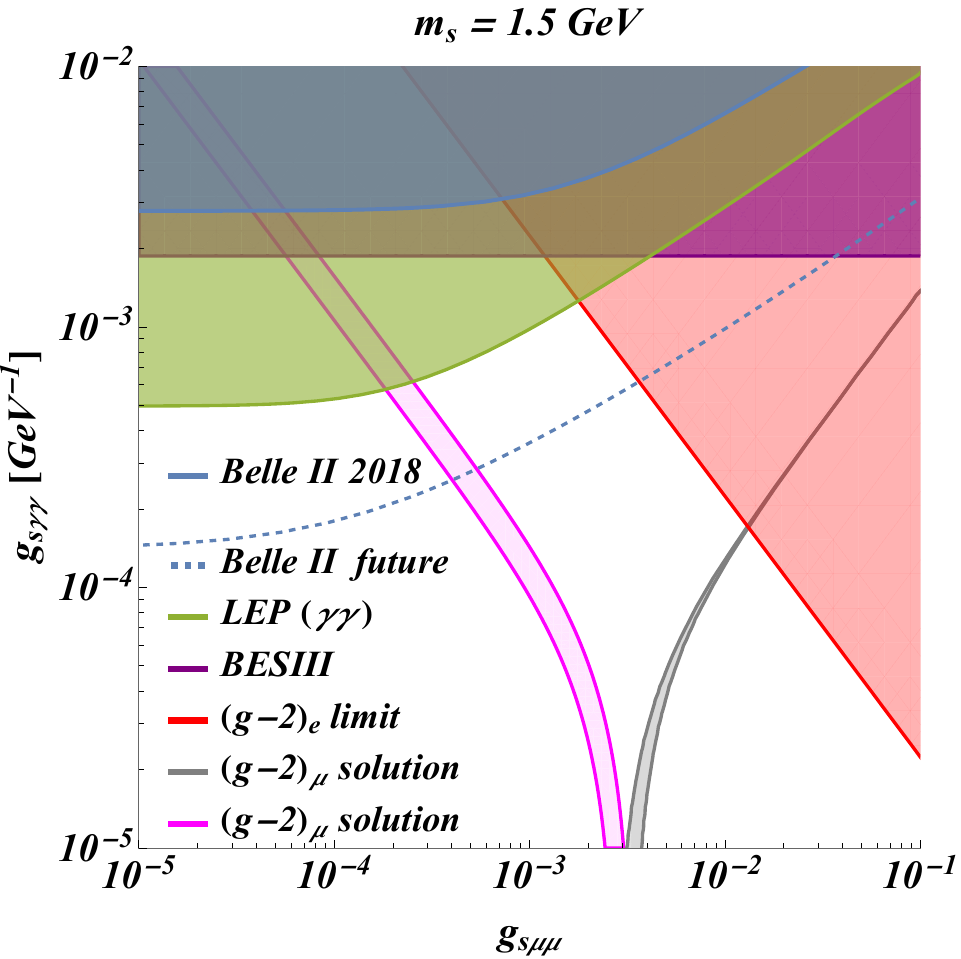} \\
\textcolor{white}{1}\\ 
\includegraphics[width=0.4\textwidth]{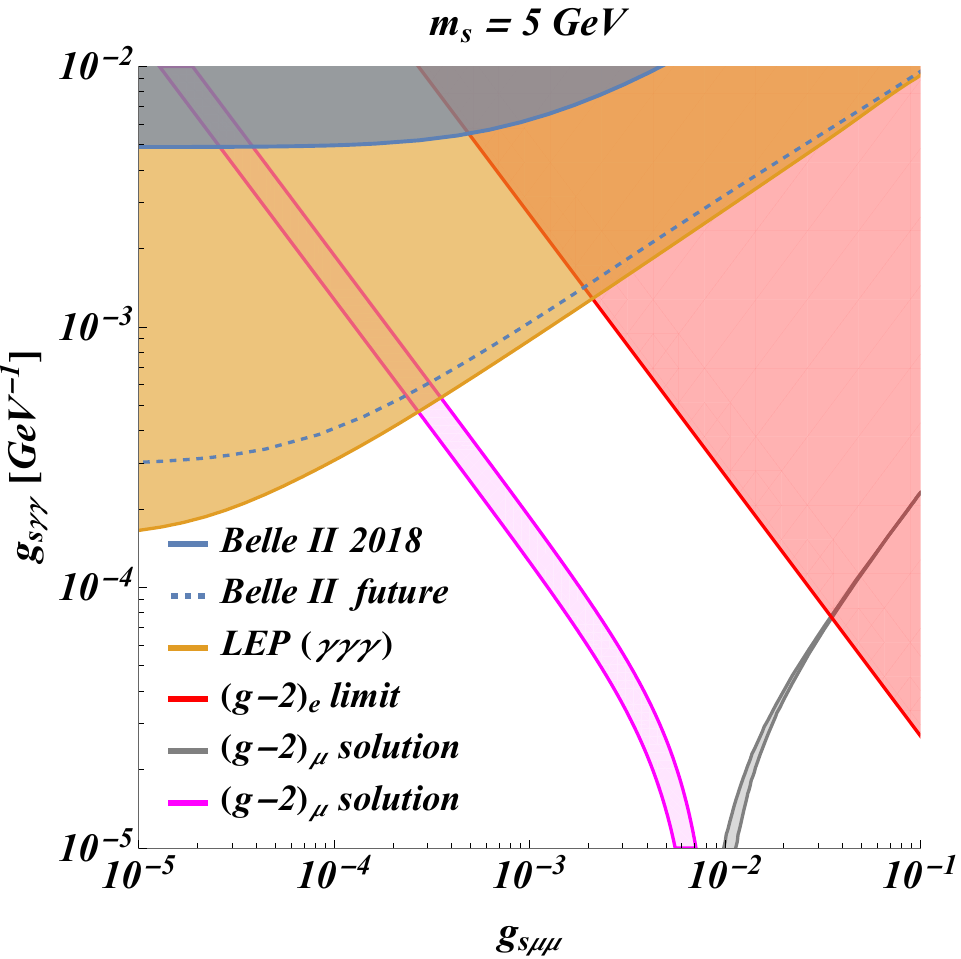}
\caption{Exclusion limits on the $\left(g_{s\mu\mu}, g_{s\gamma\gamma}\right)$ parameter space for masses $m_s = 0.4$, $1.5$ and $5$ GeV. The colored regions represent the existing bounds, the blue dashed line is the projection for the upcoming data collection at Belle II. The magenta and grey bands refer to the parameter space that could potentially resolve the $\left(g-2\right)_{\mu}$ discrepancy under the scenarios where $g_{s\gamma\gamma}$ and $g_{s\mu\mu}$ have the same and opposite signs, respectively.}
\label{fig:scalars}
\end{figure}

We also observe that the sensitivity decreases for relatively large ALP-lepton couplings, as the branching ratio $\Gamma_{a\gamma\gamma}/\Gamma_a$ becomes suppressed. Nevertheless, this regime could be potentially explored with $e^+e^-\to \tau^+\tau^-\mu^+\mu^-$ and other similar reactions, which would further constrain the parameter space and challenge the pseudoscalar explanation of $\left(g-2\right)_{\mu}$ already at the current generation of colliders and with existing data.

On the other hand, if the discrepancy in $\left(g-2\right)_{\mu}$ is eventually resolved within the Standard Model framework, the region above the grey band will be excluded, thereby utterly reducing further the parameter space available for potential ALPs.

\section{Studying scalars through $\left(g-2\right)_l$ and electron-positron colliders}\label{sec3}

Scalars in the few hundred MeV to GeV mass range represent another scenario frequently discussed in the context of New Physics~\cite{Beacham:2019nyx,Batell:2016ove,Marciano:2016yhf,Abdallah:2020biq,Winkler:2018qyg,Chen:2015vqy}. Despite the current measurements of Higgs boson properties fitting well within the Standard Model, the structure of the Higgs sector could potentially be more complicated, possibly including an additional scalar doublet~\cite{Athron:2021iuf,Berlin:2018bsc,Arcadi:2019lka}. Furthermore, particles of this nature typically have couplings to fermions that are proportional to $m_f$. Possible dark scalars are also widely discussed in different contexts and can be favored for many reasons~\cite{Mondino:2020lsc,Chen:2015vqy,Davoudiasl:2012ag,Fabbrichesi:2020wbt,GAMBIT:2017gge}.

In general, the limits for scalars are less extensively studied compared to those for pseudoscalars. The impact of scalars on the anomalous magnetic moment of the muon was discussed in~\cite{Chen:2015vqy,Marciano:2016yhf,Liu:2018xkx}. In the following, we discuss the differences between scalar and pseudoscalar scenarios, and present bounds derived from both magnetic moment measurements and $e^+e^-$ collider data.

The Lagrangian of scalar-lepton and scalar-photon interactions has some notable modifications, compared to the case of pseudoscalars. It is effectively given by

\begin{equation}
\mathcal{L}= -g_{sll} s \, \Bar{l} l - \frac{g_{s \gamma \gamma}}{4}sF^{\mu \nu}F_{\mu \nu}- \frac{g_{s \gamma Z}}{2}sF^{\mu \nu}Z_{\mu \nu}.
\end{equation}

The relation \eqref{eq:ratio} between $\gamma\gamma$ and $\gamma Z$ couplings still holds in the scalar case. One-loop corrections to $\left(g-2\right)_l$ are given by

\begin{align}
& \Delta a^Y_l = \frac{g_{sll}^2 }{8 \pi^2 } \frac{1}{r_s^2} \int_0^1  \frac{\left(1-z\right)^2\left(1+z\right)}{z + \frac{\left(1-z\right)^2}{r_s^2}} \, dz, \\
\begin{split}
& \Delta a_l^{BZ} \simeq \\ 
& \frac{ g_{sll} m_l}{8\pi^2}\ln{\Lambda^2} \left(g_{s\gamma\gamma}-\frac{4\sin^2{\theta_w}-1}{4\sin{\theta_w}\cos{\theta_w}}g_{s\gamma Z}\right).
\end{split}
\end{align}

The Yukawa contribution is positive and is capable of resolving the $\left(g-2\right)_{\mu}$ discrepancy on it's own, as explored in~\cite{Chen:2015vqy}. When the scalar-photon coupling is incorporated, two scenarios emerge. Either $g_{see}$ and $g_{s\gamma\gamma}$ have opposite signs, leading to partial cancellation between Fig. \ref{fig:dyuk} and Fig. \ref{fig:dbrz}, or they share the same sign and sum up. Both possibilities can accommodate the muon anomaly resolution, in contrast to the pseudoscalar case, where only one scenario aligns. For small values of $g_{s\gamma\gamma}$ the difference between them vanishes. 

Implementing these adjustments, we can apply the same search strategy which was used for ALPs in the previous section of this work. The partial widths of $s \to \gamma \gamma$ and $s \to l^+l^-$ decays are

\begin{align}
& \Gamma_{s \gamma \gamma} = \frac{g^2_{s \gamma \gamma}m_s^3}{64\pi} ,\\
& \Gamma_{sll} =\frac{g_{sll}^2}{\pi m_s^2}  \left(\frac{m_s^2}{4} - m_l^2\right)^{3/2}  \theta\left[m_s - 2m_l\right].
\end{align}

Fig. \ref{fig:sgmumu} shows the projection onto $\left(m_s, g_{s\gamma\gamma}\right)$ plane for the fixed value $g_{s\mu\mu}=10^{-3}$. Fig. \ref{fig:scalars} illustrates the constraints on the $\left(g_{s\mu\mu},g_{s\gamma\gamma}\right)$ parameter space for three different masses, $m_s = 0.4$, $1.5$ and $5$ GeV. Unlike the ALP case, the parameter space aligning within the $\left(g-2\right)_{\mu}$ resolution is not restricted and extends down to $g_{s\gamma\gamma} \to 0$. The rest of our conclusions regarding the pseudoscalar case also apply to the scalar case.

\section{Conclusion}\label{sec4}

In this paper we thoroughly examined ALPs and scalars coupled to leptons and photons in a minimal way. Constraints from $e^+e^- \to 3 \gamma$ annihilation and electron magnetic moment from both existing and upcoming experiments were derived, and the available parameter space for the possible $\left(g-2\right)_{\mu}$ resolution was tested.

Our analysis has emphasized the significance of simultaneous consideration of couplings of ALPs and scalars to both leptons and photons, revealing the intricate interplay between these interactions and their implications for experimental constraints. We have demonstrated the potential for narrowing down the parameter space for ALPs and scalars in the MeV to GeV mass range and conclude that while existing experimental results already impose stringent bounds on the parameter space relevant to $\left(g-2\right)_{\mu}$, a substantial portion of the parameter space remains unexplored.

There is still room for further improvement in this work. A detailed study of processes such as $e^+e^-\to \text{leptons}$ at Belle II and BESIII collider experiments  could provide a direct probe of $g_{all}$ couplings and yield more stringent bounds on them. This could potentially narrow down the parameter space further or even provide the complete coverage of the parameter space relevant to $\left(g-2\right)_{\mu}$. 

In addition to that, we assumed that studied scalars and pseudoscalars have 100$\%$ branching ratio into visible states. However, in many theoretical models such particles play the role of mediators between Standard Model particles and the dark sector. Clearly, the contribution of light dark matter particles with sub-GeV masses could modify the established constraints if the pair production threshold is exceeded. On the other hand, incorporating such particles produces a lot of uncertainties, which greatly complicates the analysis of experimental bounds.

\section*{Acknowledgments}
This work was supported by the Deutsche Forschungsgemeinschaft (DFG, German Research Foundation), in part through the Research Unit [Photon-photon interactions in the Standard Model and beyond, Projektnummer 458854507 - FOR 5327], and in part through the Cluster of Excellence [Precision Physics, Fundamental Interactions, and Structure of Matter] (PRISMA$^+$ EXC 2118/1) within the German Excellence Strategy (Project ID 39083149).

\appendix
\section{Finite parts of Barr-Zee loop contributions}\label{appendix:a}

The finite part of the Barr-Zee diagram shown in Fig. \ref{fig:dbrz} can be represented in terms of the function

\begin{equation}
\Delta = m_l^2y^2 + m_a^2\left(1-x\right)-i\varepsilon. 
\end{equation}

ALP contribution calculated with the Lagrangian \eqref{eq:der} in case of the internal photon line is given by

\begin{equation}
\begin{split}
& \text{Finite} \left[\Delta a^{BZ}_l\right] = - \frac{m_l g_{all}g_{a\gamma\gamma}}{8\pi^2}  \\
& \times \int_0^1\int_0^x   \biggl\{ \left(1+3y\right)\ln{\Delta}+\frac{m_l^2y^3}{\Delta} \biggr\} dy\,dx.
\end{split}
\end{equation}

If the internal line represents a $Z$-boson instead, we need to replace

\begin{align}
& g_{a\gamma\gamma} \to -\frac{4\sin^2{\theta_w}-1}{4\cos{\theta_w}\sin{\theta_w}} g_{a\gamma Z}, \label{eq:final1}\\
& \Delta \to \Delta + m_Z^2\left(x-y\right). \label{eq:final2}
\end{align}

We also note that the difference between this result and the one which can be obtained with the reduced Lagrangian \eqref{eq:yuk} is only a small constant term in the finite part of the Barr-Zee diagram. This term, in principle, can be absorbed by the redefinition of $\Lambda$.

In case of a scalar particle and the internal photon line we get

\begin{equation}
\text{Finite}\left[\Delta a^{BZ}_l\right] = - \frac{m_l g_{sll}g_{s\gamma\gamma}}{4\pi^2} \int_0^1\int_0^x \ln{\Delta} \, dy\,dx.
\end{equation}

The contribution of a $Z$-boson can be obtained by performing in this formula substitutions similar to \eqref{eq:final1} and \eqref{eq:final2}.

\bibliographystyle{apsrev4-1} 
\bibliography{bibliography}

\end{document}